\documentclass[a4paper,11pt]{article}

\pdfoutput=1

\usepackage{jcappub}

\usepackage[english]{babel}
\usepackage{units}
\usepackage{graphicx}
\usepackage{amsmath}
\usepackage{amssymb}
\usepackage{subfigure}
\usepackage{lineno}

\usepackage{tikz}
\usepackage{pgf}
\usepackage{3dplot} 

\definecolor{dark-gray}{gray}{0.3}

\newcommand{\uvect}[1]{\mathbf{\hat{#1}}}

\newcommand{\figwidth}{0.8}

\title{Polarized radio emission from extensive air showers measured with LOFAR}

\author[a,1]{P.~Schellart\note{Corresponding author.}}
\author[a]{S.~Buitink}
\author[a]{A.~Corstanje}
\author[a]{J.~E.~Enriquez}
\author[a,b,c,d]{H.~Falcke}
\author[a,b]{J.~R.~H\"orandel}
\author[a,e]{M.~Krause}
\author[a,b]{A.~Nelles}
\author[a]{J.~P.~Rachen}
\author[f]{O.~Scholten}
\author[a]{S.~ter Veen}
\author[a]{S.~Thoudam}
\author[f]{T.~N.~G.~Trinh}

\affiliation[a]{Department of Astrophysics/IMAPP, Radboud University Nijmegen,\\ P.O. Box 9010, 6500 GL Nijmegen, The Netherlands}
\affiliation[b]{Nikhef, Science Park Amsterdam,\\ 1098 XG Amsterdam, The Netherlands}
\affiliation[c]{Netherlands Institute for Radio Astronomy (ASTRON),\\ Postbus 2, 7990 AA Dwingeloo, The Netherlands} 
\affiliation[d]{Max-Planck-Institut f\"{u}r Radioastronomie,\\ Auf dem H\"ugel 69, 53121 Bonn, Germany}
\affiliation[e]{Deutsches Elektronen-Synchrotron (DESY),\\ Platanenallee 6, 15738 Zeuthen, Germany}
\affiliation[f]{KVI-CART, University of Groningen,\\ P.O. Box 72, 9700 AB Groningen, The Netherlands}

\emailAdd{P.Schellart@astro.ru.nl}

\abstract{We present LOFAR measurements of radio emission from extensive air showers. We find that this emission is strongly polarized, with a median degree of polarization of nearly $99\%$, and that the angle between the polarization direction of the electric field and the Lorentz force acting on the particles, depends on the observer location in the shower plane. This can be understood as a superposition of the radially polarized charge-excess emission mechanism, first proposed by Askaryan and the geomagnetic emission mechanism proposed by Kahn and Lerche. We calculate the relative strengths of both contributions, as quantified by the charge-excess fraction, for $163$ individual air showers. We find that the measured charge-excess fraction is higher for air showers arriving from closer to the zenith. Furthermore, the measured charge-excess fraction also increases with increasing observer distance from the air shower symmetry axis. The measured values range from $(3.3\pm 1.0)\%$ for very inclined air showers at $\unit[25]{m}$ to $(20.3\pm 1.3)\%$ for almost vertical showers at $\unit[225]{m}$. Both dependencies are in qualitative agreement with theoretical predictions.}

\arxivnumber{1406.1355}


\begin{document}
\maketitle
\flushbottom

\section{Introduction}
Cosmic rays impinging on the Earth's atmosphere induce showers of secondary particles. The motions of electrons and positrons in the electromagnetic part of the shower produce radio emission. Such radio emission, in the form of strong coherent nanosecond timescale pulses, has been detected \cite{Jelley:1965,Allan:1966,Falcke:2005} and studied by several experiments \cite{Huege:2012,Ardouin:2005,Kelley:2011}. It is now generally assumed that two mechanisms are principally responsible for the emission measured at MHz frequencies.

The dominant contribution to the radio emission is \emph{geomagnetic} in origin \cite{Kahn:1966,Allan:1971,Falcke:2003,Falcke:2005,Codalema2009}. Electrons and positrons in the shower are accelerated in opposite directions by the Lorentz force exerted by the Earth's magnetic field.
The radio emission generated in this manner is linearly polarized in the direction of the Lorentz force $\uvect{e}_{\vec{v}\times\vec{B}}$. Here $\vec{v}$ is the propagation velocity vector of the shower and $\vec{B}$ represents the Earth's magnetic field.

A secondary contribution to the radio emission results from the net excess of electrons at the front of the shower \cite{Askaryan:1962}. This excess is built up from electrons that are knocked out of atmospheric molecules by interactions with shower particles and by a net depletion of positrons due to annihilation. This \emph{charge-excess} contribution to the emission is also linearly polarized, but now radially with respect to the shower axis.

The resulting emission measured at ground level is the coherent sum of both components. Interference between these components may be constructive or destructive, depending on the position of the observer relative to the shower. Furthermore, the emission is strongly beamed in the forward direction due to the relativistic velocities of the particles. Additionally, the emission propagates through the atmosphere, which has a non-unity index of refraction that changes with height. This gives rise to relativistic time-compression effects most prominently resulting in a ring of amplified emission around the Cherenkov angle \cite{Vries:2011}. These time compression effects do not influence the polarization of the emission \cite{de-Vries:2013},  but do contribute to the highly asymmetric total intensity pattern measured on the ground \cite{Huege:2013}.

Recently, advances in simulations of air shower radio emission have converged to similar results \cite{Huege:2012a}. Moreover, it has been demonstrated that such simulations accurately predict the complex emission pattern and can be used to derive air shower parameters \cite{Buitink:2013}. This reflects the fact that the emission mechanisms are now well incorporated in them \cite{James:2011}. However, most models describe the emission at a microscopic level by calculating the contributions of each particle, in a full CORSIKA \cite{Heck:1998} or AIRES \cite{Sciutto:1999} shower simulation \cite{Huege:2013,Alvarez-Muniz:2012,Marin:2012}. This makes it difficult to disentangle the precise contributions of the geomagnetic and charge-excess components.

It is important to stress here that a separation into macroscopic emission mechanisms is not necessary for a correct modeling of the emission (the charge-excess and geomagnetic effects are both captured by the distributions and motions of the simulated charged particles). However, a description in the form of macroscopic emission mechanisms can contribute to a better physical understanding of the conditions within an air shower \cite{Werner:2012,Alvarez-Muniz:2014}.

In \cite{de-Vries:2013}, it is predicted that the relative contributions of the charge-excess and geomagnetic components fall off differently with increasing opening angle from the emission maximum. This is perceived as an increase of the charge-excess fraction (the ratio between the strengths of both contributions to the total electric field) with increasing radial distance from the shower axis. Moreover, for air showers arriving at larger zenith angles the bulk of the emission is generated further away from the observer and thus the same distance to the shower axis corresponds to a smaller opening angle (where the charge-excess fraction is lower). Therefore the charge-excess fraction should also decrease with increasing zenith angle.

An experimental indication of the presence of the charge-excess component was found by the CODALEMA experiment as a shift of the radio signal maximum with respect to the particle core \cite{Marin:2011}. A measurement of a radially polarized emission component, consistent with that produced by the charge-excess mechanism, was obtained by the Auger Engineering Radio Array (AERA). This contribution was further quantified for their specific sample of air showers, giving an average strength of $(14\pm2)\%$ in amplitude when compared to the geomagnetic contribution \cite{Aab:2014}. However, these measurements have a large antenna spacing and thus sample the radio emission of each shower only at a limited number of locations. This limits testing of the predicted dependence of the charge-excess fraction on the distance from the shower axis and the arrival direction of the shower.

Here we present high antenna density measurements of polarized radio emission from air showers with the LOFAR radio telescope. The instrumental setup is described in section~\ref{sec:instrumental_setup}. The measurements and data reduction techniques are described in sections \ref{sec:measurements} and \ref{sec:reconstructing_polarized_radio_emission}. Section~\ref{sec:procedure} describes the analysis procedure followed by a discussion of statistical and systematic uncertainties in section~\ref{sec:uncertainties}. Results are presented in section~\ref{sec:results}, and the article concludes in section \ref{sec:conclusions}.
\section{Instrumental setup}
\label{sec:instrumental_setup}
The Low Frequency Array (LOFAR) \cite{van-Haarlem:2013}, is a digital radio telescope, observing in the frequency range $\unit[10-240]{MHz}$. It was designed as a flexible instrument, capable of carrying out multiple modes of observation simultaneously. One of these modes is the measurement of radio emission from extensive air showers \cite{Schellart:2013}. For this purpose LOFAR is equipped with ring buffers, that store the raw voltage traces of each individual antenna in the array for up to $\unit[5]{s}$. When a trigger is received the ring buffers are frozen and their contents copied over the network to a central storage location. This trigger can either be generated by inspecting the signal with the on-board FPGA\footnote{Field Programmable Gate Array.} or received over the network from an external source.

For the purpose of cosmic-ray measurements a trigger is currently generated by a dedicated particle detector array. The LOFAR Radboud Air shower Array (LORA) \cite{Thoudam:2014}, consists of 20 scintillator detectors. A trigger is sent to LOFAR when an air shower is detected with an energy exceeding $E\gtrsim\unit[2\cdot 10^{16}]{eV}$. The corresponding radio emission for showers of this energy is at the lower limit of detectability by LOFAR for favorable shower geometries.

The particle detector array is located in the core of LOFAR. Here, the highest density of radio antennas is found, which makes the setup ideal for cosmic-ray detections. LOFAR consists of two types of antennas, the Low Band Antennas (LBAs) covering the frequency range $\unit[10-90]{MHz}$ and the High Band Antennas (HBAs) for $\unit[110-240]{MHz}$ (excluding the highly polluted $\unit[90-110]{MHz}$ commercial FM-radio band). Radio emission from cosmic rays has been measured in both frequency bands \cite{Schellart:2013,Nelles:2014}. The analysis presented here focusses on the LBA measurements. 

\begin{figure}
\centering
\includegraphics[width=\figwidth\textwidth]{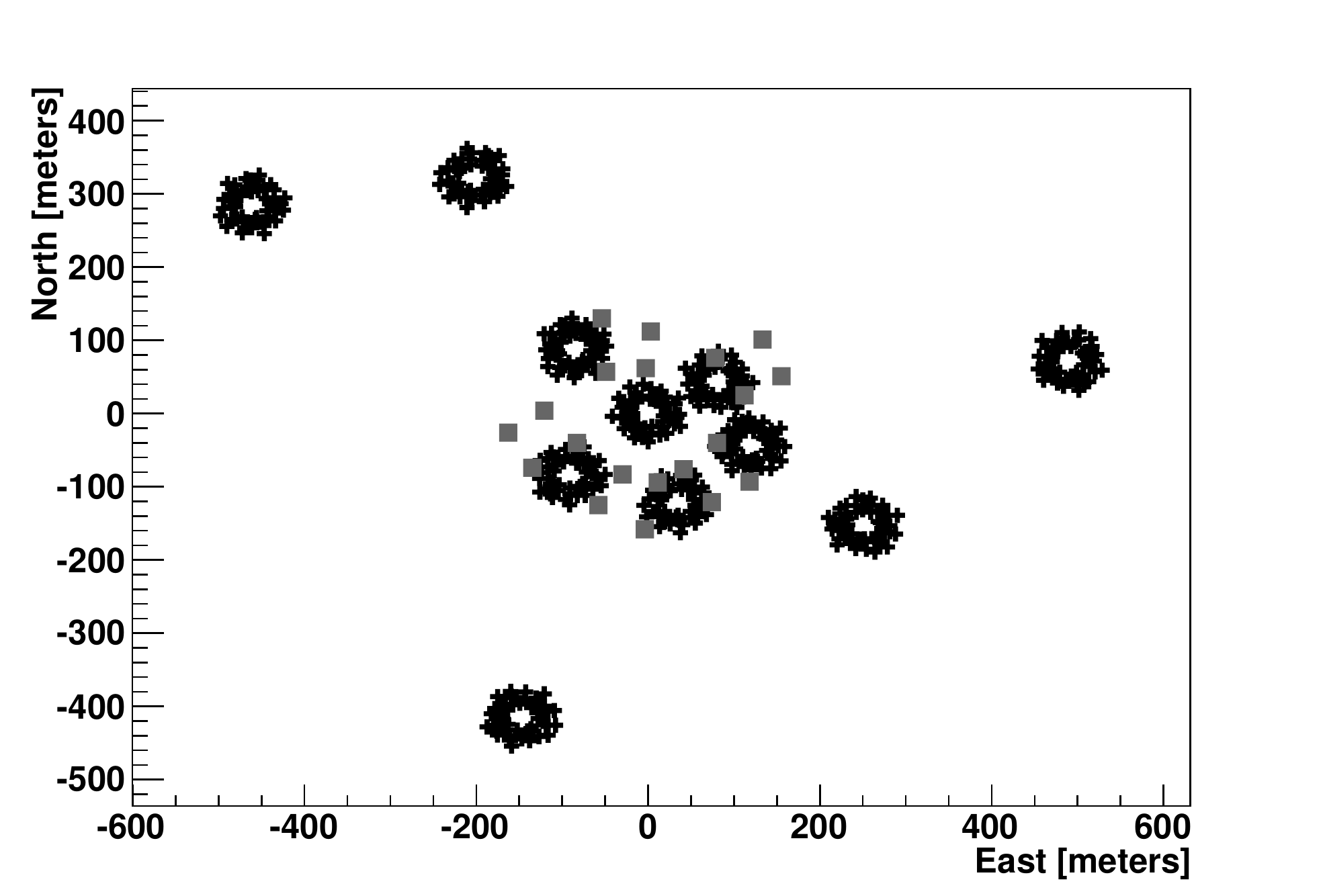}
\caption{Layout of the LOFAR stations used for this analysis. The black plus symbols represent the outer rings of low-band-antennas. The grey boxes represent the LORA particle detectors.}
\label{fig:layout}
\end{figure}

Each LBA consists of two inverted V-shaped dipoles labeled X and Y. These are aligned along the southwest to northeast and southeast to northwest direction respectively. They are grouped into stations, which in the core consist of 96 antennas each. Due to signal path limitations, only half of these antennas can be active during any given observation and, usually either the inner circle or the outer ring is selected. The $\sim\unit[2]{km}$ diameter LOFAR core consists of 24 such stations with the highest density offered by the six stations located in a $\unit[350]{m}$ diameter region. An overview of the layout of all stations used in this analysis is given in figure~\ref{fig:layout}.

\section{Measurements}
\label{sec:measurements}
For this analysis, data collected with the outer rings of low-band antennas between June 2011 and January 2014 are used. The inner rings of low-band antennas are excluded for this analysis as crosstalk effects are present in some of the more closely spaced antennas, which are currently not included in the antenna simulations needed for polarization analysis. All data corresponding to a LORA trigger are stored together for offline analysis and constitute an event.

In the offline analysis these data are processed per station as described in \cite{Schellart:2013}. The most important steps, for polarization analysis, are briefly summarized here.

First radio frequency interference (RFI) is identified and removed. Thereafter the received power in each dipole, outside of the cosmic-ray signal, is dominated by Galactic emission. This is used to perform a relative gain calibration between all dipoles, independently for the X and Y dipoles. An absolute calibration is currently not yet available at LOFAR.

An initial estimate for the arrival direction of the air shower is given by the particle detector array. The predicted geometric signal arrival time delays for this direction are used to coherently add the signals from all dipoles in one station, thus forming a beam in this direction. This increases the signal-to-noise ratio for a cosmic-ray signal from that direction by approximately a factor of seven. When a pulse is found with a signal-to-noise ratio exceeding three the station is marked for further processing.

In the next processing step the arrival direction of the air shower is reconstructed from a plane wave fit to the arrival times of the pulse maxima (defined as the maximum of its Hilbert envelope). In this analysis only air showers where four or more stations have a successful direction reconstruction are included.

\begin{figure}
\centering
\includegraphics[width=\figwidth\textwidth]{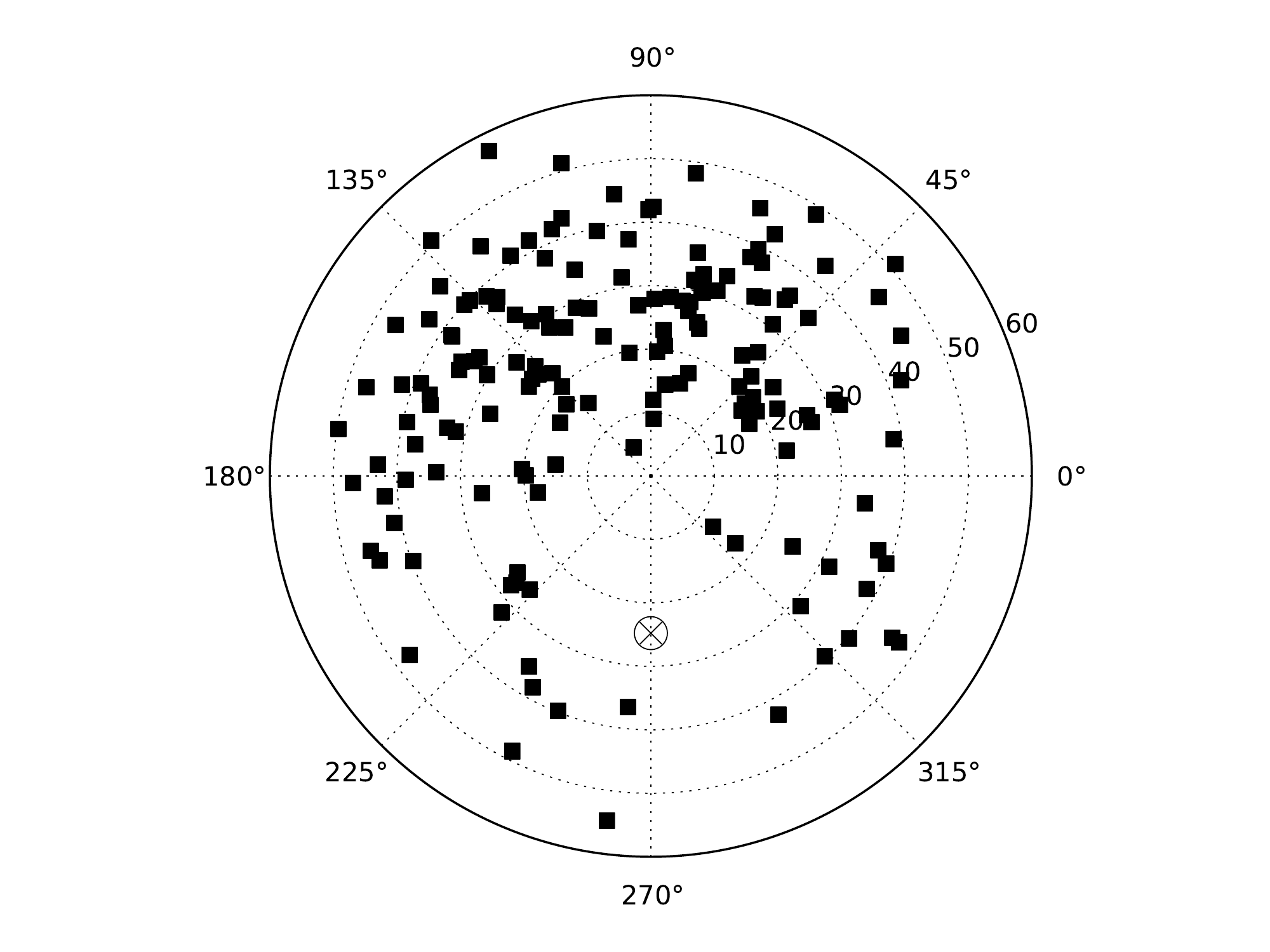}
\caption{Reconstructed arrival directions for all air showers used in this analysis (squares). Also indicated (cross) is the direction of the magnetic field ($\unit[18.6]{\mu T}$ North, $\unit[45.6]{\mu T}$ downward) at LOFAR.}
\label{fig:arrival_directions}
\end{figure}

There are 206 air showers that meet this criterium. An overview of their reconstructed arrival directions is given in figure~\ref{fig:arrival_directions}. As can be seen in this figure, the arrival directions of the measured air showers are not isotropically distributed. This is a well known selection effect caused by the dependence of the dominant geomagnetic emission component on the geomagnetic angle $\alpha$ between the arrival direction of the air shower and the Earth's magnetic field \cite{Wilson:1971,Codalema2009,Schellart:2013}. This makes showers arriving at greater geomagnetic angles easier to detect and hence more numerous in a given sample.

Of these 206 showers a further 18 are excluded from this analysis, because they are likely influenced by thunderstorm conditions. These are defined by lightning strikes being recorded by the Royal Netherlands Meteorological Institute (KNMI) within two hours of the event in the vicinity of the LOFAR core. Five events do not coincide with a thunderstorm but show a similar polarization pattern and are also excluded. An additional 18 events were too weak to get a reliable estimate of the shower core position (see section~\ref{sec:shower_plane}). Two additional events were excluded because the polarization reconstruction failed for a single station. This leaves a total of 163 air showers used for this analysis, for which the emission reaching the ground is sampled at between 192 and 528 distinct locations.

\section{Reconstructing polarized radio emission}
\label{sec:reconstructing_polarized_radio_emission}
\begin{figure}
\centering
\tdplotsetmaincoords{60}{120}
\pgfmathsetmacro{\rvec}{0.8}
\pgfmathsetmacro{\thetavec}{30}
\pgfmathsetmacro{\phivec}{60}
\begin{tikzpicture}[scale=4,tdplot_main_coords]
\coordinate (O) at (0,0,0);
\tdplotsetcoord{P}{\rvec}{\thetavec}{\phivec}
\draw[thick,->] (-1,0,0) -- (1,0,0) node[anchor=north east]{East, x};
\draw[thick,->] (0.4,0.4,0) -- (-0.4,-0.4,0) node[anchor=south]{X};
\draw[thick,->] (0,-0.8,0) -- (0,1,0) node[anchor=north west]{North, y};
\draw[thick,->] (-0.4,0.4,0) -- (0.4,-0.4,0) node[anchor=east]{Y};
\draw[thick,->] (0,0,0) -- (0,0,1) node[anchor=south]{Zenith, z};
\draw[-stealth,-] (O) -- (P) node[anchor=south west]{};
\draw[dashed] (O) -- (Pxy);
\draw[dashed] (P) -- (Pxy);
\tdplotdrawarc{(O)}{0.2}{0}{\phivec}{}{}
\node at (25:.3){$\phi$};
\tdplotsetthetaplanecoords{\phivec}
\tdplotdrawarc[tdplot_rotated_coords]{(0,0,0)}{0.3}{0}{\thetavec}{anchor = north}{}
\node[tdplot_rotated_coords] at (15:0.4){$\theta$};
\draw[dashed,tdplot_rotated_coords,color=dark-gray] (\rvec,0,0) arc (0:90:\rvec);
\draw[dashed,color=dark-gray] (\rvec,0,0) arc (0:360:\rvec);
\tdplotsetrotatedcoords{\phivec}{\thetavec}{0}
\tdplotsetrotatedcoordsorigin{(P)}
\draw[thick,tdplot_rotated_coords,->] (0,0,0) -- (.2,0,0) node[anchor=north west]{$\uvect{e}_{\theta}$};
\draw[thick,tdplot_rotated_coords,->] (0,0,0) -- (0,.2,0) node[anchor=west]{$\uvect{e}_{\phi}$};
\draw[thick,tdplot_rotated_coords,->] (0,0,0) -- (0,0,.3) node[anchor=west]{-$\uvect{e}_{v}$};
\foreach \angle in {0,0}
{
\coordinate (P) at (0,0,\sintheta);
\tdplotsetthetaplanecoords{\angle}
\tdplotdrawarc[dashed,tdplot_rotated_coords,color=dark-gray]{(O)}{\rvec}{-90}{90}{}{}
}
\end{tikzpicture}
\caption[Geometry of coordinate systems]{On-sky polarization coordinate frame ($\uvect{e}_{\theta}$, $\uvect{e}_{\phi}$, $-\uvect{e}_{\vec{v}}$). Also depicted is the (north, east, zenith) coordinate frame corresponding to the $x$, $y$ and $z$-axis, respectively. Furthermore the dipole antennas $X$ and $Y$ are shown (projected onto the ground plane).}
\label{fig:project}
\end{figure}
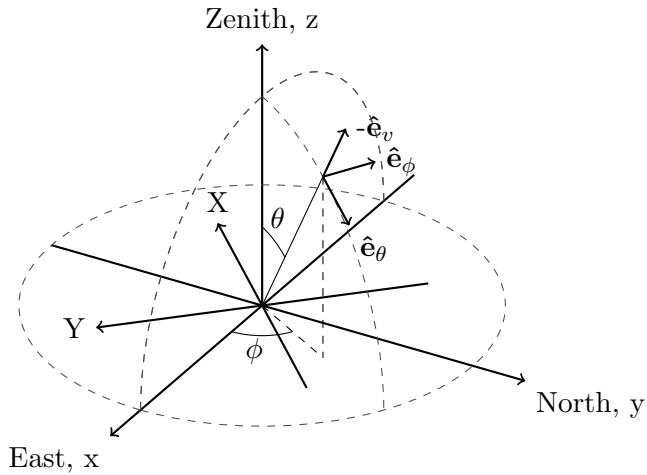

Using the reconstructed direction of the air shower, the full time traces measured by the $X$ and $Y$ dipoles are combined, while correcting for their frequency-dependent complex gain \cite{Schellart:2013}. These gains are obtained from antenna simulations and are calculated for a plane wave, arriving from direction $-\uvect{e}_{v}$, where $\vec{v}$ is the propagation velocity vector of the air shower, and polarized along $\uvect{e}_{\theta}$ or $\uvect{e}_{\phi}$ as defined in figure~\ref{fig:project}. The resulting combined signals are thus the electric field components along $\uvect{e}_{\theta}$ and $\uvect{e}_{\phi}$.

For the present analysis these are subsequently projected, following figure~\ref{fig:coordinates}, onto the $\uvect{e}_{\vec{v}\times\vec{B}}$ and $\uvect{e}_{\vec{v}\times\vec{v}\times\vec{B}}$ directions. Here $\vec{B}$ again represents the geomagnetic field.
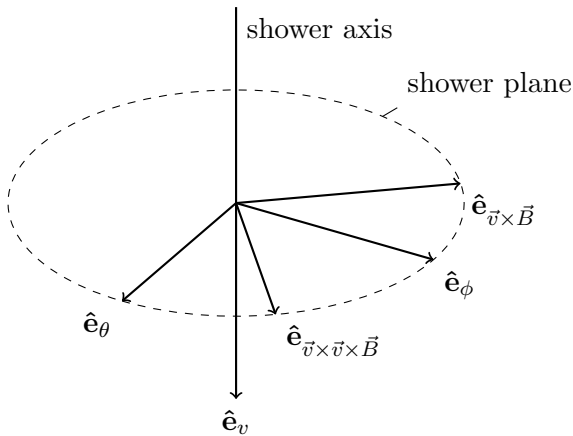
\begin{figure}
\centering
\tdplotsetmaincoords{60}{120}

\begin{tikzpicture}[scale=3.0,tdplot_main_coords]
\coordinate (O) at (0,0,0);
\draw[thick,->] (0:0) -- (0:1.0) node[anchor=north east]{$\uvect{e}_{\theta}$};
\draw[thick,->] (90:0) -- (90:1.0) node[anchor=north west]{$\uvect{e}_{\phi}$};
\draw[thick,->] (0,0,0) -- (0,0,-1.0) node[anchor=north]{$\uvect{e}_{v}$};
\draw[thick,-] (0,0,0) -- (0,0,1.0) node[anchor=north west]{shower axis};
\draw[thin,-] (170:1.0) -- (170:1.1) node[anchor=south west]{shower plane};
\draw[thick,->] (40:0.0) -- (40:1.0) node[anchor=north west]{$\uvect{e}_{\vec{v}\times\vec{v}\times\vec{B}}$};
\draw[thick,->] (40+90:0.0) -- (40+90:1.0) node[anchor=north west]{$\uvect{e}_{\vec{v}\times\vec{B}}$};
\tdplotdrawarc[dashed]{(0,0,0)}{1.0}{0}{360}{}{}
\end{tikzpicture}
\caption[coordinates]{The natural coordinate frame for air shower polarization measurements is given by the unit vectors $\uvect{e}_{\vec{v}}$, $\uvect{e}_{\vec{v}\times\vec{B}}$ and $\uvect{e}_{\vec{v}\times\vec{v}\times\vec{B}}$. The polarization components of the emission as reconstructed by the antenna model $\uvect{e}_{\theta}$, $\uvect{e}_{\phi}$ are transformed to this frame by simple rotation around the shower axis.}
\label{fig:coordinates}
\end{figure}

Note that for this procedure it is assumed that the emission has no component along the propagation direction $\uvect{e}_{v}$ \cite{Schellart:2013}.

\subsection{Observer positions in the shower plane}
\label{sec:shower_plane}
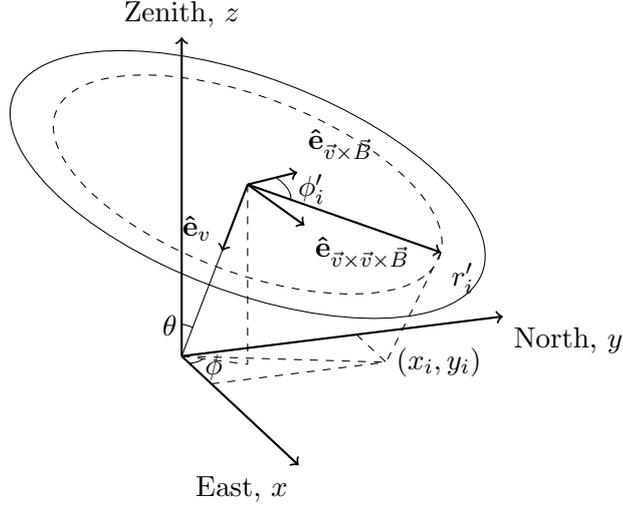
\begin{figure}
\centering

\tdplotsetmaincoords{70}{70}
\pgfmathsetmacro{\rvec}{0.9}
\pgfmathsetmacro{\thetavec}{20}
\pgfmathsetmacro{\phivec}{50}
\pgfmathsetmacro{\radiusveclen}{0.9}
\pgfmathsetmacro{\radiusvecangle}{15}
\pgfmathsetmacro{\vxbvecangle}{-20}
\pgfmathsetmacro{\vxbveclen}{0.35}

\begin{tikzpicture}[scale=3.0,tdplot_main_coords]
\coordinate (O) at (0,0,0);
\tdplotsetcoord{P}{\rvec}{\thetavec}{\phivec};
\tdplotsetcoord{L}{cos(\thetavec * pi / 180) * \radiusveclen * \radiusveclen / sqrt((\radiusveclen * cos((\phivec + \radiusvecangle) * pi / 180))^2 + (cos(\thetavec * pi / 180) * \radiusveclen * sin((\phivec + \radiusvecangle) * pi / 180))^2)}{90}{(\phivec + \radiusvecangle)};
\draw[dashed] (O) -- (L);
\draw[thick,->] (0,0,0) -- (1.5,0,0) node[anchor=north east]{East, $x$};
\draw[thick,->] (0,0,0) -- (0,1.5,0) node[anchor=north west]{North, $y$};
\draw[thick,->] (0,0,0) -- (0,0,1.5) node[anchor=south]{Zenith, $z$};
\draw[-stealth,-] (O) -- (P) node[anchor=south west]{};
\draw[dashed] (O) -- (Pxy);
\draw[dashed] (P) -- (Pxy);
\draw[dashed] (Lx) -- (Lxy);
\draw[dashed] (Ly) -- (Lxy) node[anchor=west]{$(x_i,y_i)$};
\tdplotdrawarc{(O)}{0.1}{0}{\phivec}{}{}
\node at (25:.2){$\phi$};
\tdplotsetthetaplanecoords{\phivec}
\tdplotdrawarc[tdplot_rotated_coords]{(0,0,0)}{0.15}{\thetavec}{0}{anchor = east}{$\theta$}
\tdplotsetrotatedcoords{\phivec}{\thetavec}{0}
\tdplotsetrotatedcoordsorigin{(P)}
\tdplotdrawarc[tdplot_rotated_coords]{(0,0,0)}{1.1}{0}{360}{}{}
\draw[thick,tdplot_rotated_coords,->] (\radiusvecangle:0) -- (\radiusvecangle:\radiusveclen) node[anchor=north west]{$r_i'$};
\draw[thick,tdplot_rotated_coords,->] (\vxbvecangle:0) -- (\vxbvecangle:\vxbveclen) node[anchor=north west]{$\uvect{e}_{\vec{v}\times\vec{v}\times\vec{B}}$};
\draw[thick,tdplot_rotated_coords,->] (\vxbvecangle+90:0) -- (\vxbvecangle+90:\vxbveclen) node[anchor=south west]{$\uvect{e}_{\vec{v}\times\vec{B}}$};
\tdplotdrawarc[tdplot_rotated_coords]{(0,0,0)}{0.2}{\vxbvecangle+90}{\radiusvecangle}{anchor=west}{$\phi_i'$};
\draw[dashed,tdplot_rotated_coords] (\radiusvecangle:\radiusveclen) -- (Lxy);
\tdplotdrawarc[tdplot_rotated_coords, dashed]{(0,0,0)}{\radiusveclen}{0}{360}{}{}
\draw[thick,tdplot_rotated_coords,->] (0,0,0) -- (0,0,.-\vxbveclen) node[anchor=south east]{$\uvect{e}_{v}$};
\end{tikzpicture}
\caption[Shower plane geometry]{Geometry of the shower plane for a shower arriving with azimuth and zenith angles $\phi$ and $\theta$ respectively. The direction of the shower plane is defined by its normal vector $\uvect{e}_{v}$. All antenna positions $(x_i, y_i)$ are projected onto this plane giving $r_i$, the distance to the shower axis and $\phi'$ the angle with the $\uvect{e}_{\vec{v}\times\vec{B}}$ direction. Note that the direction of $\uvect{e}_{\vec{v}\times\vec{B}}$ in this figure is chosen for clarity in display and does not accurately reflect the direction of the magnetic field at LOFAR.}
\label{fig:shower_plane}
\end{figure}

Given a position for the shower core and arrival direction, the antenna positions can be projected onto the shower plane as given by figure~\ref{fig:shower_plane}. Here each antenna $i$ is represented by the polar coordinates $\phi_{i}'$ measured from the $\uvect{e}_{\vec{v}\times\vec{B}}$ axis (positive in the direction of the $\uvect{e}_{\vec{v}\times\vec{v}\times\vec{B}}$ axis), and $r_{i}'$ the distance from the shower axis.

While the particle detector array offers an initial estimate of the core position this is not reliable when the shower core is not contained within the particle detector array, as is often the case for the measured air showers. Therefore, the shower core position is determined by fitting a two-dimensional lateral distribution function to the integrated pulse power of the radio signal, following the procedure described in \cite{Nelles:2015}. This provides the core position with an estimated statistical uncertainty of $\sim\unit[15]{m}$.

The arrival direction is the average of those obtained per station. This has an estimated statistical uncertainty of $\sim 1^{\circ}$. A better angular resolution of $\sim 0.1^{\circ}$ can be obtained by fitting a hyperbolic wavefront to the arrival times at the antennas for all stations \cite{Corstanje:2015}, however this is not needed for the current analysis.

\subsection{Stokes parameters}
Due to the expected elliptically polarized nature of the received signal \cite{Schoorlemmer:2008} it is difficult to directly compare the electric field amplitudes in both polarization components. A better approach is to use time integrated quantities such as the Stokes parameters \cite{Jackson:1975}:
\begin{align}
I &= \frac{1}{n}\sum_{i=0}^{n-1}(E_{i,\vec{v}\times\vec{B}}^{2} + \hat{E}_{i,\vec{v}\times\vec{B}}^{2} + E_{i,\vec{v}\times\vec{v}\times\vec{B}}^{2} + \hat{E}_{i,\vec{v}\times\vec{v}\times\vec{B}}^{2})\label{eq:stokes_I},\\
Q &= \frac{1}{n}\sum_{i=0}^{n-1}(E_{i,\vec{v}\times\vec{B}}^{2} + \hat{E}_{i,\vec{v}\times\vec{B}}^{2} - E_{i,\vec{v}\times\vec{v}\times\vec{B}}^{2} - \hat{E}_{i,\vec{v}\times\vec{v}\times\vec{B}}^{2})\label{eq:stokes_Q},\\
U &= \frac{2}{n}\sum_{i=0}^{n-1}(E_{i,\vec{v}\times\vec{B}}E_{i,\vec{v}\times\vec{v}\times\vec{B}} + \hat{E}_{i,\vec{v}\times\vec{B}}\hat{E}_{i,\vec{v}\times\vec{v}\times\vec{B}})\label{eq:stokes_U},\\
V &= \frac{2}{n}\sum_{i=0}^{n-1}(\hat{E}_{i,\vec{v}\times\vec{B}}E_{i,\vec{v}\times\vec{v}\times\vec{B}} - E_{i,\vec{v}\times\vec{B}}\hat{E}_{i,\vec{v}\times\vec{v}\times\vec{B}})\label{eq:stokes_V}.
\end{align}
The integration is performed over $n=5$ samples, of $\unit[5]{ns}$ each, centered around the pulse maximum. Here $E_{i,j}$ is sample $i$ of the $j-$th electric field component and $\hat{E}_{i,j}$ its Hilbert transform.

For an elliptically polarized signal one can calculate from the Stokes parameters the angle that the semi-major axis of the polarization ellipse makes with the $\uvect{e}_{\vec{v}\times\vec{B}}$ axis
\begin{equation}
\psi = \frac{1}{2} \tan^{-1}\left(\frac{U}{Q}\right).
\label{eq:polarization_angle}
\end{equation}
Additionally the \emph{degree of polarization} is calculated which is defined to be the fraction of the power in the polarized component of the wave
\begin{equation}
p = \frac{\sqrt{Q^2 + U^2 + V^2}}{I}.
\label{eq:degree_of_polarization}
\end{equation}

\section{Determining the contributions of the emission mechanisms}
\label{sec:procedure}
Following \cite{Aab:2014} the expected electric field at any given time $t$ can be written as
\begin{equation}
\begin{split}
\vec{E}(t) &= \vec{E}_{\mathrm{G}}(t) + \vec{E}_{\mathrm{C}}(t)\\
&= (|\vec{E}_{\mathrm{G}}(t)| + |\vec{E}_{\mathrm{C}}(t)|\cos\phi')\uvect{e}_{\vec{v}\times\vec{B}} +\\
&\quad (|\vec{E}_{\mathrm{C}}(t)|\sin\phi')\uvect{e}_{\vec{v}\times\vec{v}\times\vec{B}}.
\end{split}
\end{equation}
Here $\vec{E}_{\mathrm{G}}(t)$ is the electric field produced by the geomagnetic contribution that is purely polarized along the direction of the Lorentz force, and $\vec{E}_{\mathrm{C}}(t)$ the radial electric field, produced by charge-excess.
The charge-excess fraction is then defined as
\begin{equation}
\label{eq:charge_excess_fraction}
a\equiv \sin\alpha \frac{|E_{\mathrm{C}}|}{|E_{\mathrm{G}}|},
\end{equation}
where $|E_{\mathrm{C}}|$ is the amplitude of the electric field produced by charge-excess when the total electric field vector amplitude reaches its maximum value.

The $\sin\alpha$ factor, with $\alpha$ the angle between the magnetic field $\vec{B}$ and the propagation direction of the shower $\vec{v}$, reflects the known dependence of the geomagnetic contribution.

Thus, the electric field at the time of the pulse maximum can be written as
\begin{equation}
\vec{E}=|E_{\mathrm{G}}|\left[\left(1+\frac{a}{\sin\alpha}\cos\phi'\right)\uvect{e}_{\vec{v}\times\vec{B}} + \left(\frac{a}{\sin\alpha}\sin\phi'\right)\uvect{e}_{\vec{v}\times\vec{v}\times\vec{B}}\right].
\label{eq:electric_field_direction}
\end{equation}
Therefore, the expected angle that the electric field vector at the time of the pulse maximum makes with the $\uvect{e}_{\vec{v}\times\vec{B}}$ axis is given by
\begin{equation}
\label{eq:polarization_angle_a}
\psi' = \tan^{-1}\left(\frac{\sin\phi'}{\frac{\sin\alpha}{a} + \cos\phi'}\right).
\end{equation}
This angle is equal to the angle of the semi major axis $\psi'=\psi$ of the polarization ellipse. Thus, the charge-excess fraction, $a$, can be determined by fitting eq.~\eqref{eq:polarization_angle_a} to $\psi$ as a function of azimuthal angle $\phi'$ in the shower plane.

\section{Uncertainties}
\label{sec:uncertainties}
When determining the charge-excess fraction by fitting eq.~\eqref{eq:polarization_angle_a} to data, three main factors contribute to the uncertainty. These are background noise, through $\psi$, and the uncertainties on the reconstructed position and arrival direction of the air shower, through $\phi'$, $\psi$ and $\alpha$. Here the statistical and systematic uncertainties associated with these contributions are discussed and quantified.
\subsection{Statistical uncertainty on the angle of polarization}
\label{sec:uncertainty_angle}
Both direction dependent effects and background noise contribute to the uncertainty on the angle of polarization $\psi$.

Any uncertainty on the direction of the shower axis translates into an uncertainty on the Stokes parameters through the combination of the measured signals in both dipoles in the antenna model. In order to estimate the uncertainty on $\psi$ due to this effect a Monte Carlo procedure is employed. In this procedure the Stokes parameters and $\psi$ are determined $200$ times for each event while picking a random direction and core position from within the $1^\circ$ and $\unit[15]{m}$ uncertainties respectively. The spread (standard deviation) of the thus determined distribution of $\psi$ gives the uncertainty due to the directional dependence $\sigma_{\psi,d}$.

A second contribution to the total uncertainty on $\psi$ results from the influence of background noise. An uncertainty $\sigma_{\psi,n}$ is assigned to each measurement based on the relation
\begin{equation}
\sigma_{\psi,n} = \frac{0.37}{S/N}
\label{eq:snr_uncertainty_relation}
\end{equation}
\label{eq:signal_to_noise_definition}
where $S/N$ is the signal-to-noise ratio defined as
\begin{equation}
S/N = \sqrt{\frac{P_\mathrm{s}}{P_\mathrm{n}}}
\end{equation}
with $P_\mathrm{s}$ the total integrated power contained in the pulse and $P_\mathrm{n}$ the total integrated power in a noise window (scaled to the same duration).

To derive this relation a similar Monte Carlo procedure is used. For $200$ trials per event the Stokes parameters and $\psi$ are determined after adding background noise to the signal. The spread on the thus determined distribution of $\psi$ gives the uncertainty resulting from noise $\sigma_{\psi,n}$.

\begin{figure}
\centering
\includegraphics[width=\figwidth\textwidth]{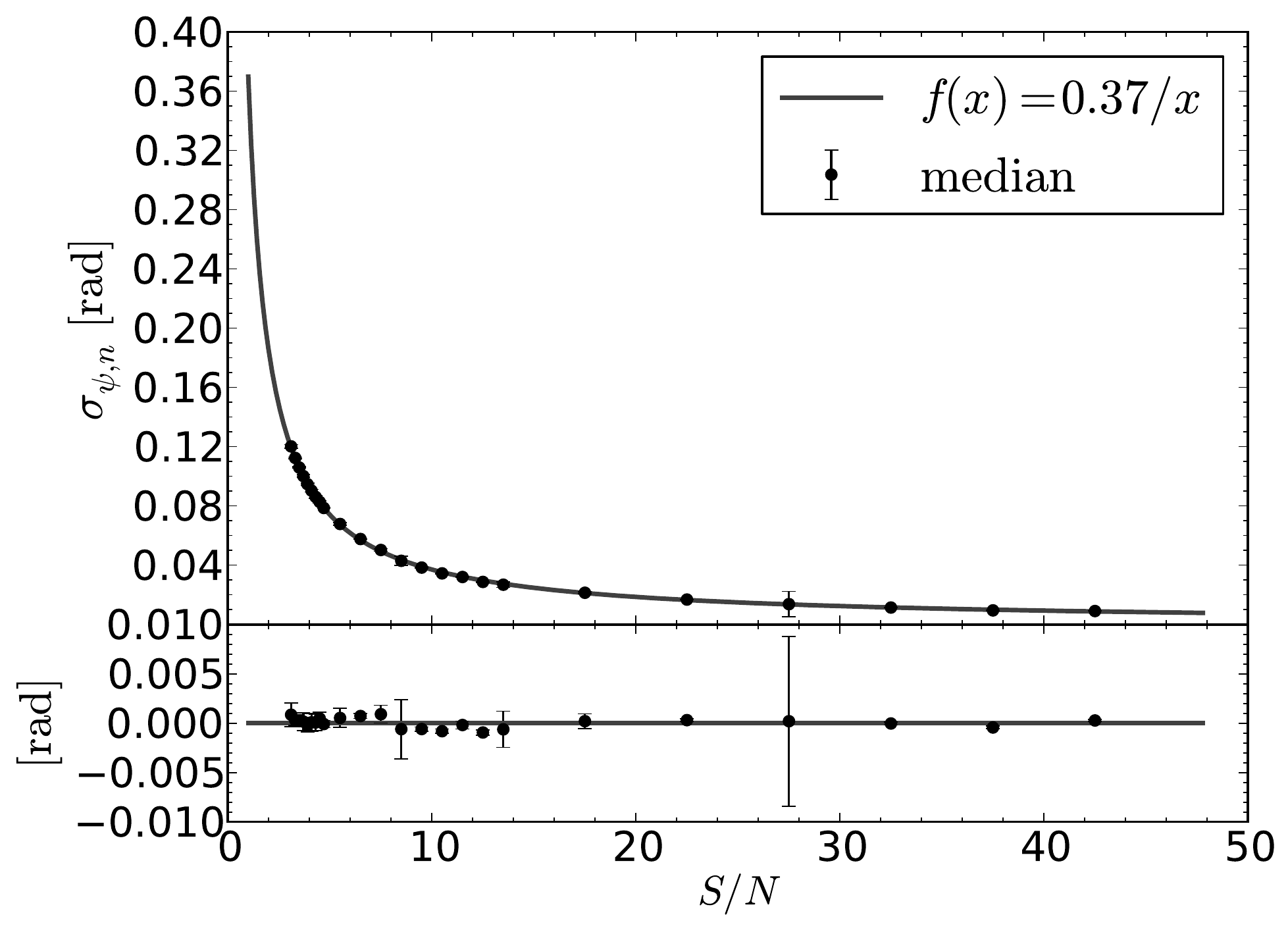}
\caption{Uncertainty contribution to the polarization angle due to background noise as a function of signal-to-noise ratio (as defined by eq.~\eqref{eq:signal_to_noise_definition}).}
\label{fig:snr_sigma_psi_n}
\end{figure}

However, because this Monte Carlo procedure lowers the signal-to-noise ratio before calculating the Stokes parameters, the resulting uncertainty contribution is overestimated (it is the contribution for $S/(\sqrt{2}N)$ instead of $S/N$) and cannot be assigned to the measurement directly. Therefore, as a first step $\sigma_{\psi,n}$ is calculated for all measurements and plotted as a function of the actual $S/N$ (corrected by a factor $\sqrt{2}$ to account for the addition of noise) in figure~\ref{fig:snr_sigma_psi_n}. In order to minimize the influence of outliers the median is calculated for each $S/N$ bin. The bin size is $0.2$ below $S/N=5$ and increases to $1$ and $5$ for higher signal-to-noise ratios to ensure a sufficient number of measurements per bin. As can be seen, the uncertainty is inversely proportional to the signal-to-noise ratio and fitting for the proportionality constant gives relation \eqref{eq:snr_uncertainty_relation}.

The total uncertainty on $\psi$ is now given by
\begin{equation}
\label{eq:uncertainty_psi}
\sigma_{\psi}=\sigma_{\psi,d} + \sigma_{\psi,n}.
\end{equation}
Because the uncertainties are in principle correlated, they are combined linearly which corresponds to the most conservative case of maximum correlation.

\subsection{Statistical uncertainty on the charge-excess fraction}
\label{sec:uncertainty_charge_excess}
As discussed in section~\ref{sec:uncertainty_angle} the uncertainty on the polarization angle $\psi$ is caused by two effects. The influence of noise, and the propagation of the uncertainties on the arrival direction and core position of the shower through the antenna model. The uncertainty on $\psi$ propagates into the uncertainty on the charge-excess fraction $a$, determined by fitting eq.~\eqref{eq:polarization_angle_a}, as $\sigma_{a,\psi}$ which is readily available from the covariance matrix of the fit. However there is a second effect. Any uncertainty on the core position and arrival direction of the shower also introduces an uncertainty on the azimuth of the observer position in the shower plane $\phi'$. This effect can be particularly strong for antennas close to the shower axis and needs to be taken into account in the uncertainty on $a$. In addition the geomagnetic angle $\alpha$ also changes for different arrival directions and hence is affected by any uncertainty on the reconstruction which propagates into $a$ through eq.~\eqref{eq:polarization_angle_a}.

In order to estimate the uncertainty on the charge-excess fraction determined from a fit of eq.~\eqref{eq:polarization_angle_a} to a set of points $(\psi_{i}, \phi'_{i})$ a Monte Carlo procedure was employed. At each step, $j$, a core position and arrival direction were selected randomly within their respective uncertainties around the measured value giving $\phi'_{i,j}$ and $\alpha_{j}$. Subsequently $a$ was determined by fitting eq.~\eqref{eq:polarization_angle_a} with fixed weights $1/\sigma_\psi$ from eq.~\eqref{eq:uncertainty_psi}. This procedure was repeated $200$ times. The standard deviation of the thus determined distribution of $a$ is taken to be the uncertainty contribution $\sigma_{a,\phi',\alpha}$ of the uncertainties on core position and arrival direction through $\phi'$ and $\alpha$ to $a$.

The total uncertainty on $a$ is found by adding the two contributions
\begin{equation}
\sigma_a = \sigma_{a,\psi} + \sigma_{a,\phi',\alpha}.
\label{eq:uncertainty_charge_excess}
\end{equation}
As in eq.~\eqref{eq:uncertainty_psi}, the combination is done linearly.

\subsection{Systematic uncertainties}
\label{sec:systematic_uncertainties}
The charge-excess fraction, $a$, is determined by fitting eq.~\eqref{eq:polarization_angle_a} to the measured angle of polarization $\psi$ as a function of angle in the shower plane $\phi'$. For the typically small values $(a\approx 0.11)$ found in measurements (see section~\ref{sec:relative_strength}), $\psi$ varies approximately sinusoidally and $a$ is related to the amplitude of the variation. Thus, any bias on $\psi$ that simply acts to rotate the measured polarization vector equally for all antennas should not influence $a$.

Such a bias on $\psi$ might be the result of inaccuracies in the antenna model. Given that the antenna model gives an excellent agreement between LOFAR data and the intensity pattern predicted by air shower simulations (see \cite{Buitink:2013}) we expect that this effect will be small. Nevertheless, the influence on $\psi$ can be quantified by adding a global offset, $\Delta\psi$, as an additional free parameter to eq.~\eqref{eq:polarization_angle_a}. It was found that $\Delta\psi\approx -0.6^\circ \pm 2.6^\circ$ on average, with the higher values obtained for showers where the $\phi'$ coverage is less uniform, and that the measured value of $a$ is not significantly altered by including $\Delta\psi$ for all air showers in this analysis. Because the deviation from a plane wavefront is small, approximately $1^\circ$ at a distance of $\unit[250]{m}$ from the shower axis \cite{Corstanje:2015}, and hence all antennas are `seeing' the shower from the same direction (with the same antenna gain), we do not expect any dependence of $\psi$ on $\phi'$ to result from the antenna model.

A second possible bias on $\psi$ may occur when the levels of background noise in the reconstructed polarization components, $\vec{E}_{\vec{v}\times\vec{B}}$ and $\vec{E}_{\vec{v}\times\vec{v}\times\vec{B}}$, are not equal \cite{Schoorlemmer:2008}. While this in principle can be corrected for by subtracting the Stokes parameters calculated on background noise alone before calculating the angle of polarization, this has the downside of increasing the statistical uncertainty. For this reason it was opted to not correct for the systematic effect and instead minimize its influence by choosing a narrow $\unit[25]{ns}$ pulse window.

To quantify the remaining systematic contribution to the measured charge-excess fraction the following procedure was followed. From an existing set of air shower simulations generated with the CoREAS \cite{Huege:2013} plugin to CORSIKA \cite{Heck:1998} a single radio pulse was arbitrarily selected. This pulse was subsequently aligned with $\uvect{e}_{\vec{v}\times\vec{B}}$ and hence $\psi=0$ by construction. For each measured air shower and each antenna the noise levels in the two reconstructed polarization components were determined in an off-pulse data block. Gaussian noise was then added to the simulated pulse consistent with the measured noise level difference. The polarization angle $\psi$ was determined for each antenna and the charge-excess, $a$, was found by fitting eq.~\eqref{eq:polarization_angle_a}. The median value of the thus determined systematic uncertainty on $a$ is very small $0.001$ and can for most purposes be neglected.

Thus, in absence of any obvious systematic effects on $\psi$ that also depend on $\phi'$ we expect systematic effects on $a$ to be low, and certainly not more than $\sim10\%$.
\section{Results}
\label{sec:results}
\begin{figure}
\centering
\includegraphics[width=\figwidth\textwidth]{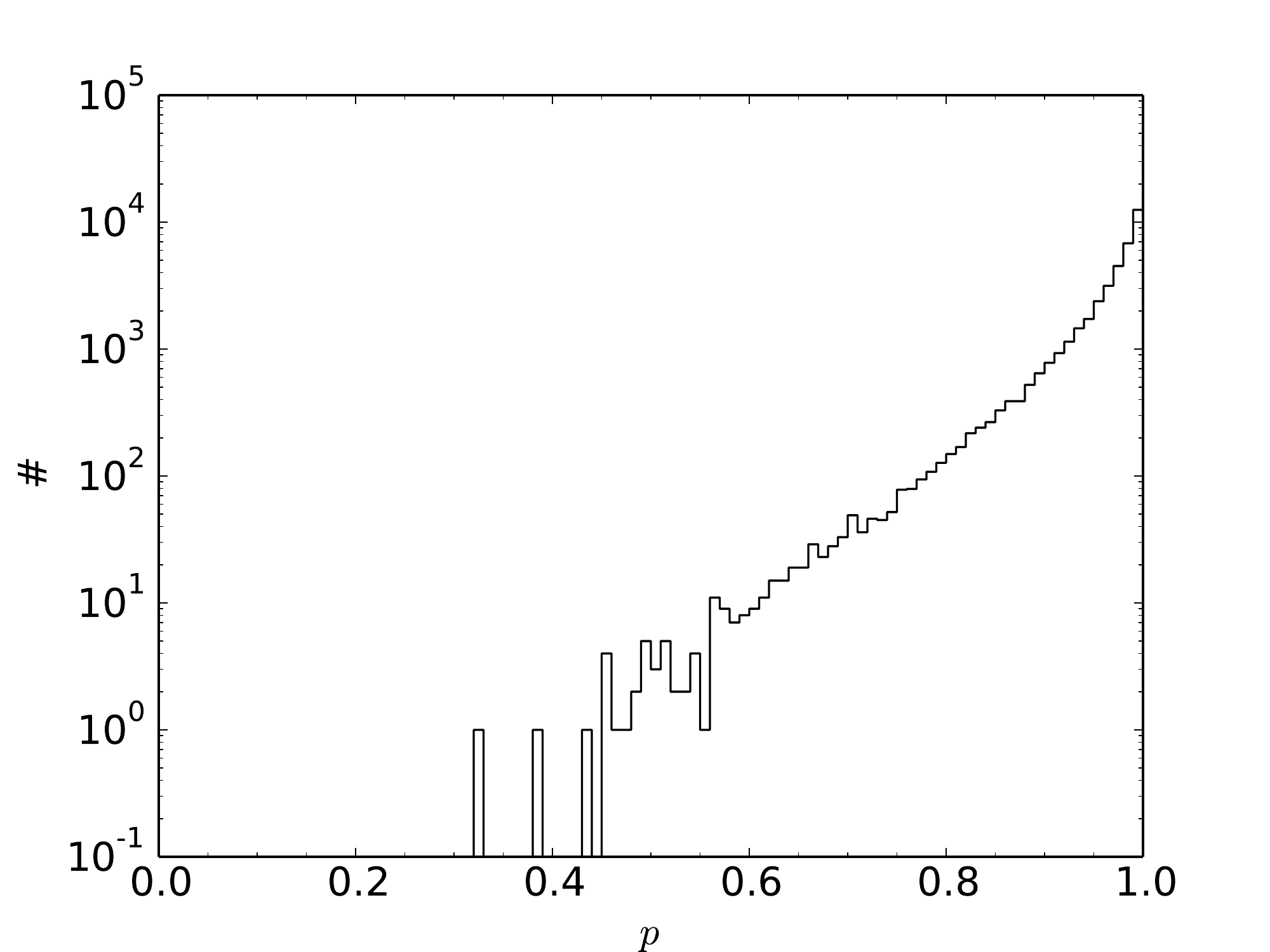}
\includegraphics[width=\figwidth\textwidth]{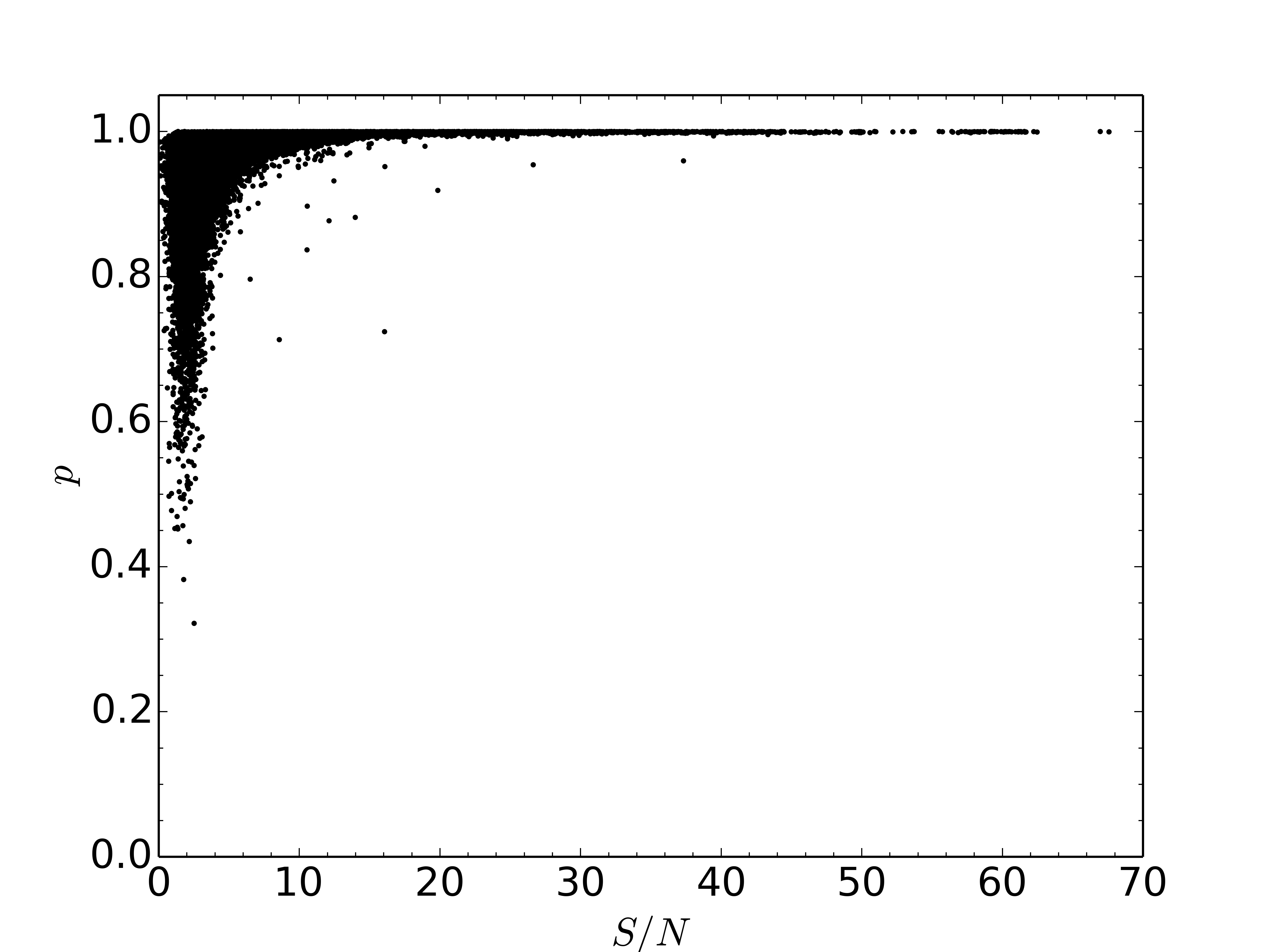}
\caption{Number of antennas showing a degree of polarization of $p$ where $p$ has been binned in intervals of 0.01 for all measured air showers (top panel) and $p$ as a function of the signal-to-noise ratio (bottom panel). The signal-to-noise ratio is defined as in eq.~\eqref{eq:signal_to_noise_definition}.}
\label{fig:degree-of-polarization}
\end{figure}

Since we expect the air shower signal to be completely polarized, the degree of polarization should be close to unity. The distribution of $p$ for all measured showers can be seen in the top panel of figure~\ref{fig:degree-of-polarization}. With a  degree of polarization of $97.3\%$, $98.9\%$ and $99.6\%$ in the $25\%$, $50\%$ (median) and $75\%$ quartiles respectively, the air shower signal is strongly polarized. For those cases where the ratio is less than unity the difference is expected to be caused by the unpolarized background noise. This can be seen in the bottom panel of figure~\ref{fig:degree-of-polarization} where the degree of polarization approaches unity for antennas with a high measured signal-to-noise ratio.

\begin{figure}
\centering
\includegraphics[width=\figwidth\textwidth]{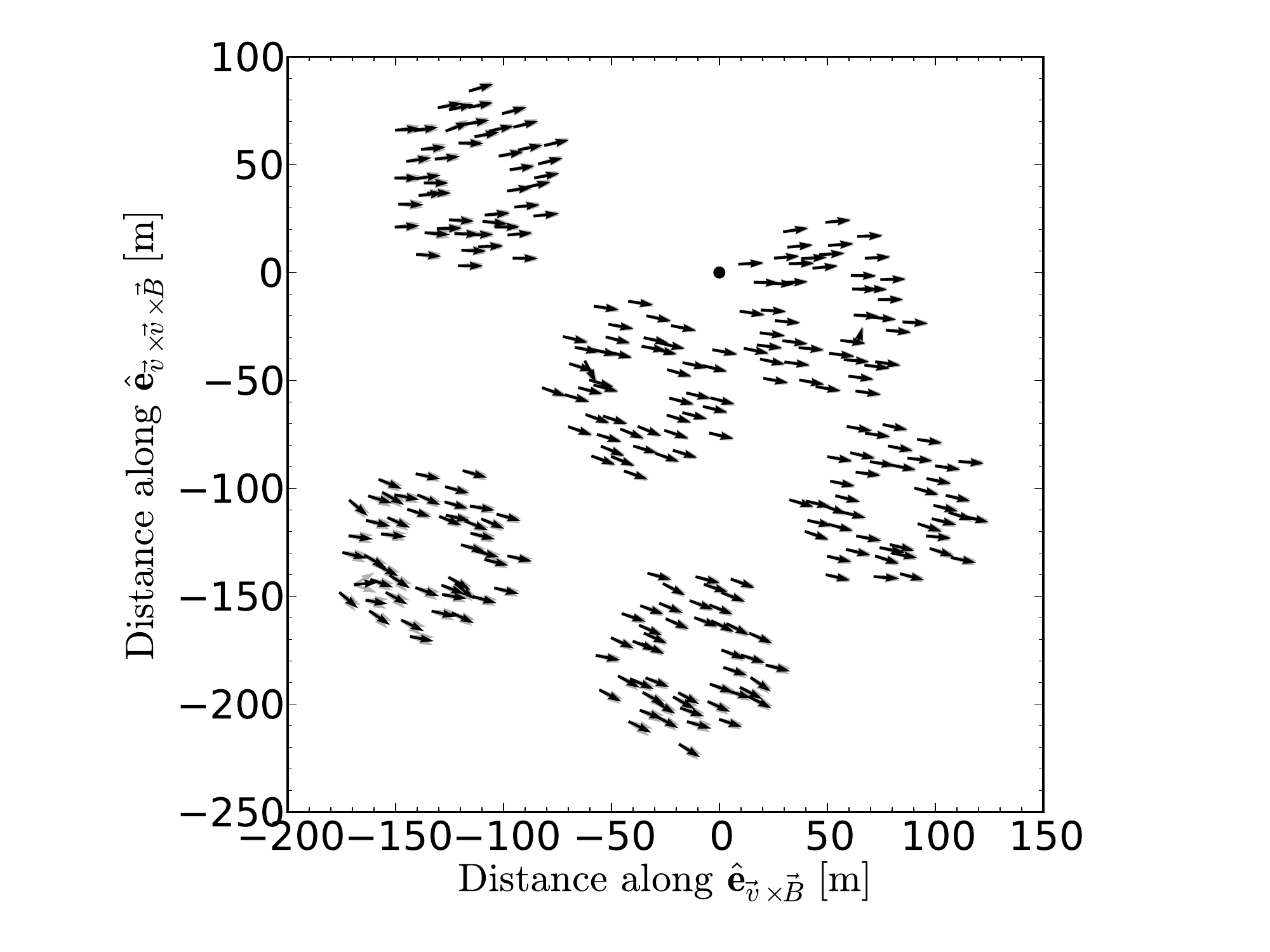}
\caption{Polarization footprint of a single air shower, as recorded with the LOFAR low-band antennas, projected onto the shower plane. Each arrow represents the electric field measured by one antenna. The direction of the arrow is defined by the polarization angle $\psi$ with the $\uvect{e}_{\vec{v}\times\vec{B}}$ axis and its length is proportional to the degree of polarization $p$. The shower axis is located at the origin (indicated by the black dot). The median uncertainty on the angle of polarization is $4^\circ$ and the value for each antenna is indicated by the grey arrows in the background. Except for a few antennas in the lower left station they are mostly small, indicating that the pattern is not the result of a random fluctuation.}
\label{fig:polarization_footprint}
\end{figure}

An instructive way to represent the polarization information is in the form of the \emph{polarization footprint} of the shower as given in figure~\ref{fig:polarization_footprint} for an example event. Here, the angle and degree of polarization are depicted as arrows for each antenna position projected in the shower plane. As expected for geomagnetically dominated emission, the arrows are roughly aligned along the $\uvect{e}_{\vec{v}\times\vec{B}}$ direction. However, a small deviation from this direction can be seen. This deviation is interpreted as being due to the charge-excess component and causes the arrows to point upward or downward, above or below the projected shower core.

\subsection{Measurement of the radially polarized emission component}
\label{sec:relative_strength}
\begin{figure}
\centering
\includegraphics[width=\figwidth\textwidth]{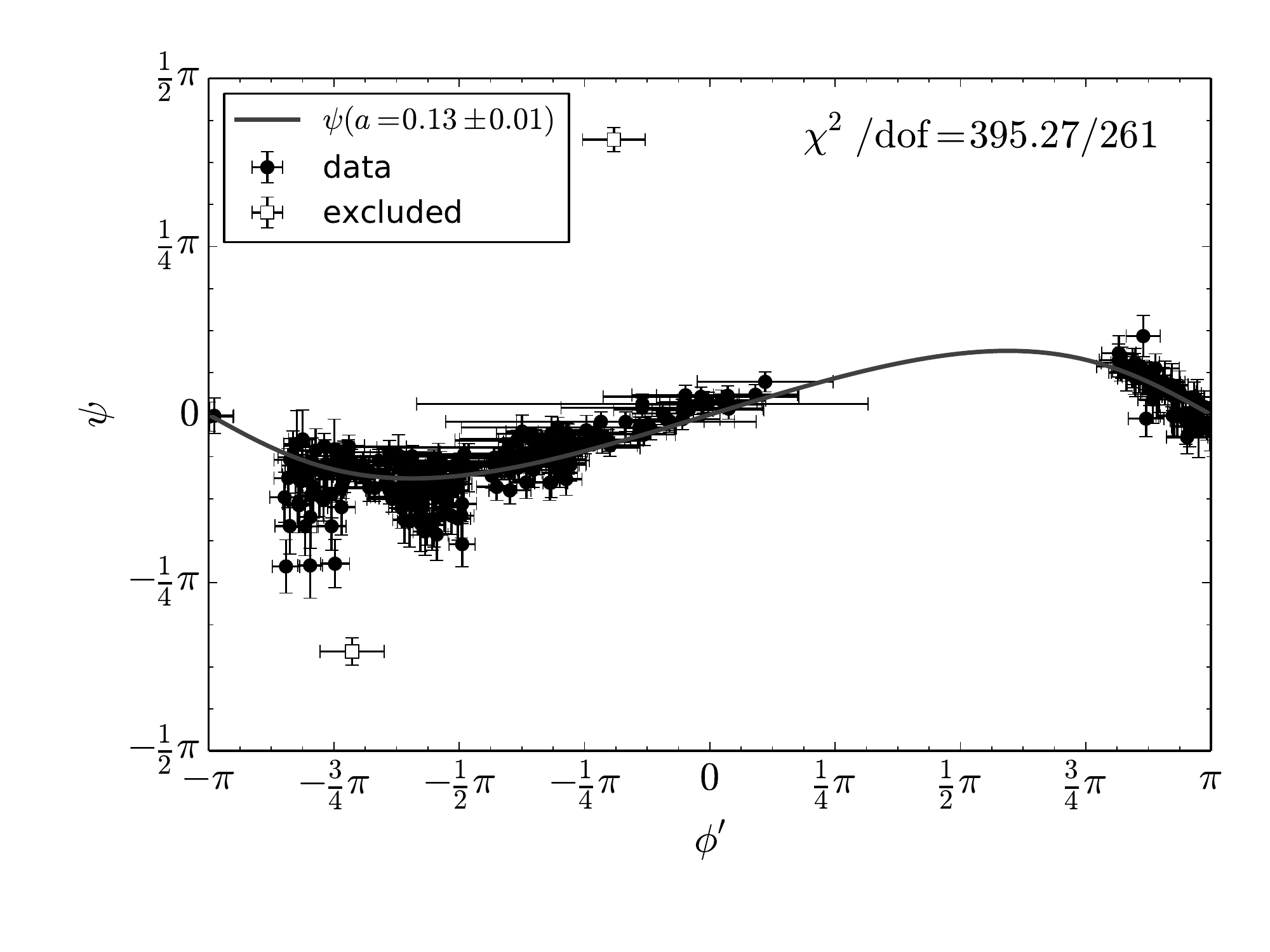}
\includegraphics[width=\figwidth\textwidth]{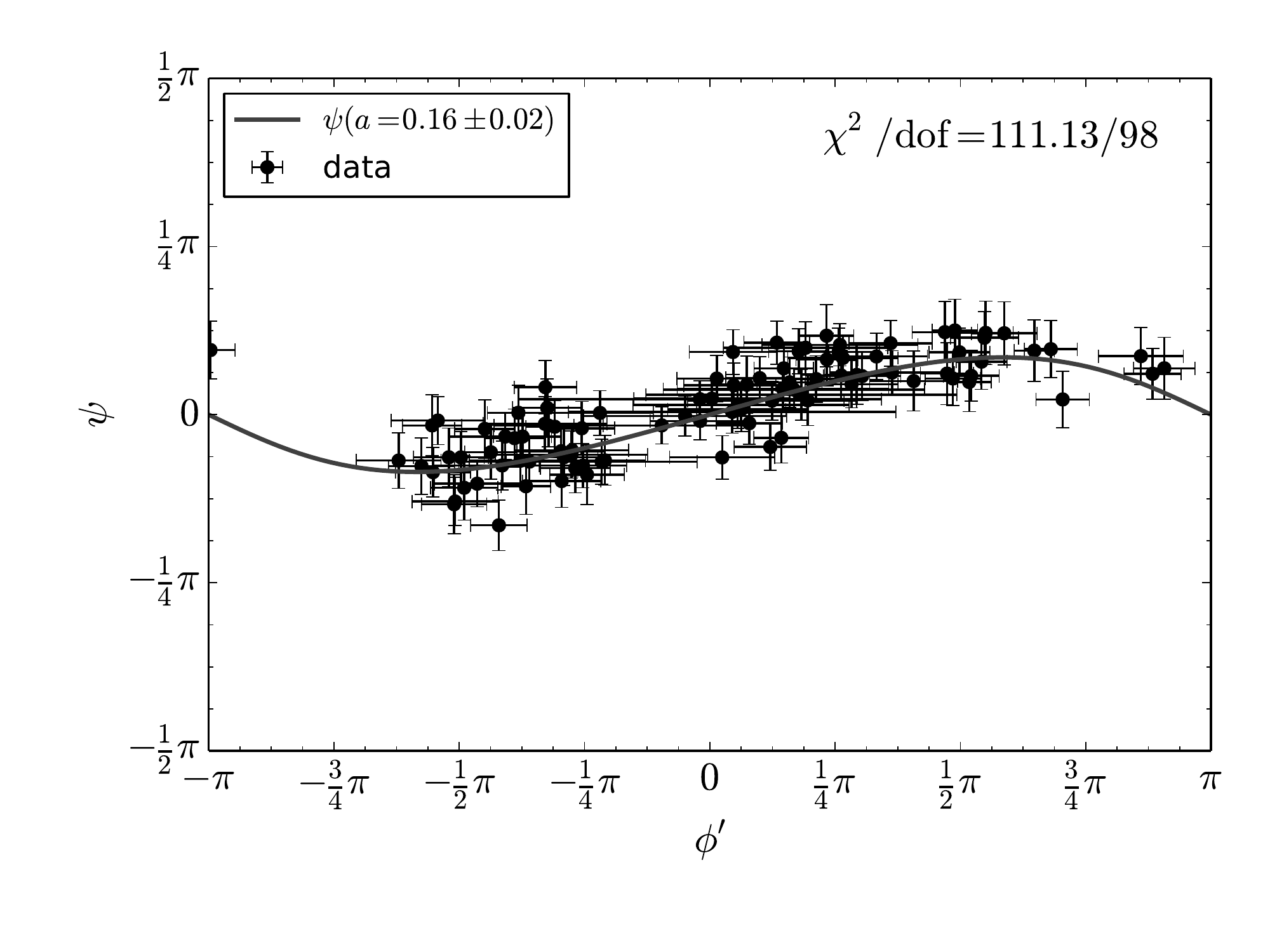}
\caption{Polarization angle $\psi$ as a function of azimuth in the shower plane $\phi'$ for two individual air showers measured with the LOFAR low-band antennas at $\unit[10-90]{MHz}$ (circles). Uncertainties on the polarization angle $\psi$ are calculated as discussed in section~\ref{sec:uncertainty_angle} and, uncertainties on the charge-excess fraction $a$ are determined as described in section~\ref{sec:uncertainty_charge_excess}. The solid line represents the function \eqref{eq:polarization_angle_a} for the best fitting value of the charge-excess fraction $a$. In the top panel two antennas are excluded from the fit (open squares), since their signals deviate by more than $10\sigma_i$ from the best fit of eq.~\eqref{eq:polarization_angle_a} in the first iteration. At maximum two percent of the antennas are allowed to be excluded.}
\label{fig:single_event_a_fit}
\end{figure}

In the presence of a radially polarized emission component, with strength $a/\sin\alpha$ relative to the geomagnetic component, we expect that the angle of polarization depends on the observer location in the shower plane according to eq.~\eqref{eq:polarization_angle_a}. In figure~\ref{fig:single_event_a_fit} this dependence can clearly be seen for two measured air showers.
\clearpage

\begin{figure}
\centering
\includegraphics[width=\figwidth\textwidth]{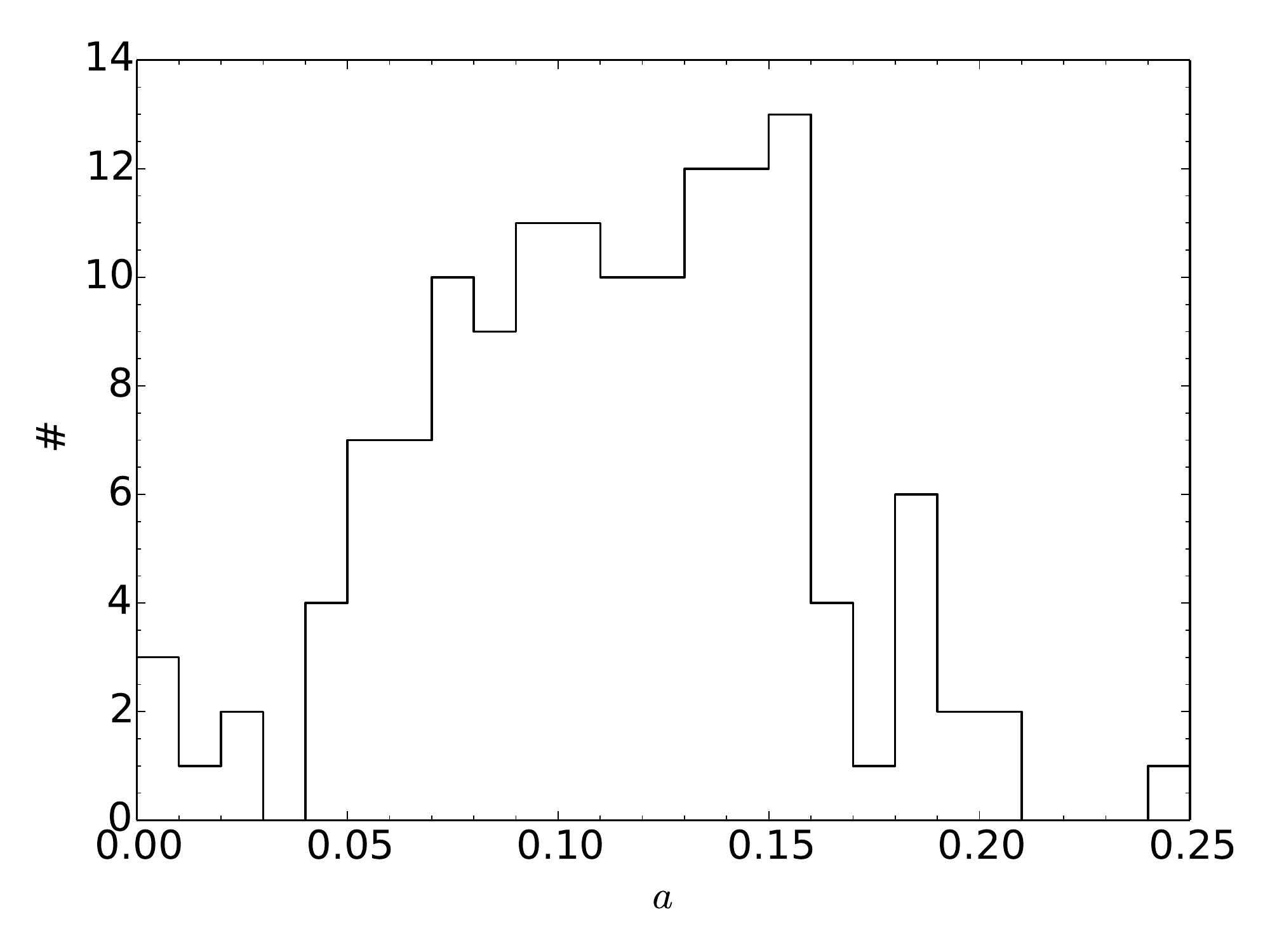}
\caption{Distribution of the charge-excess fraction, $a$, as determined from individual air shower measurements binned in intervals of $0.01$ for all measured air showers.}
\label{fig:distribution_single_event_a_fit}
\end{figure}

\begin{figure}
\centering
\includegraphics[width=\figwidth\textwidth]{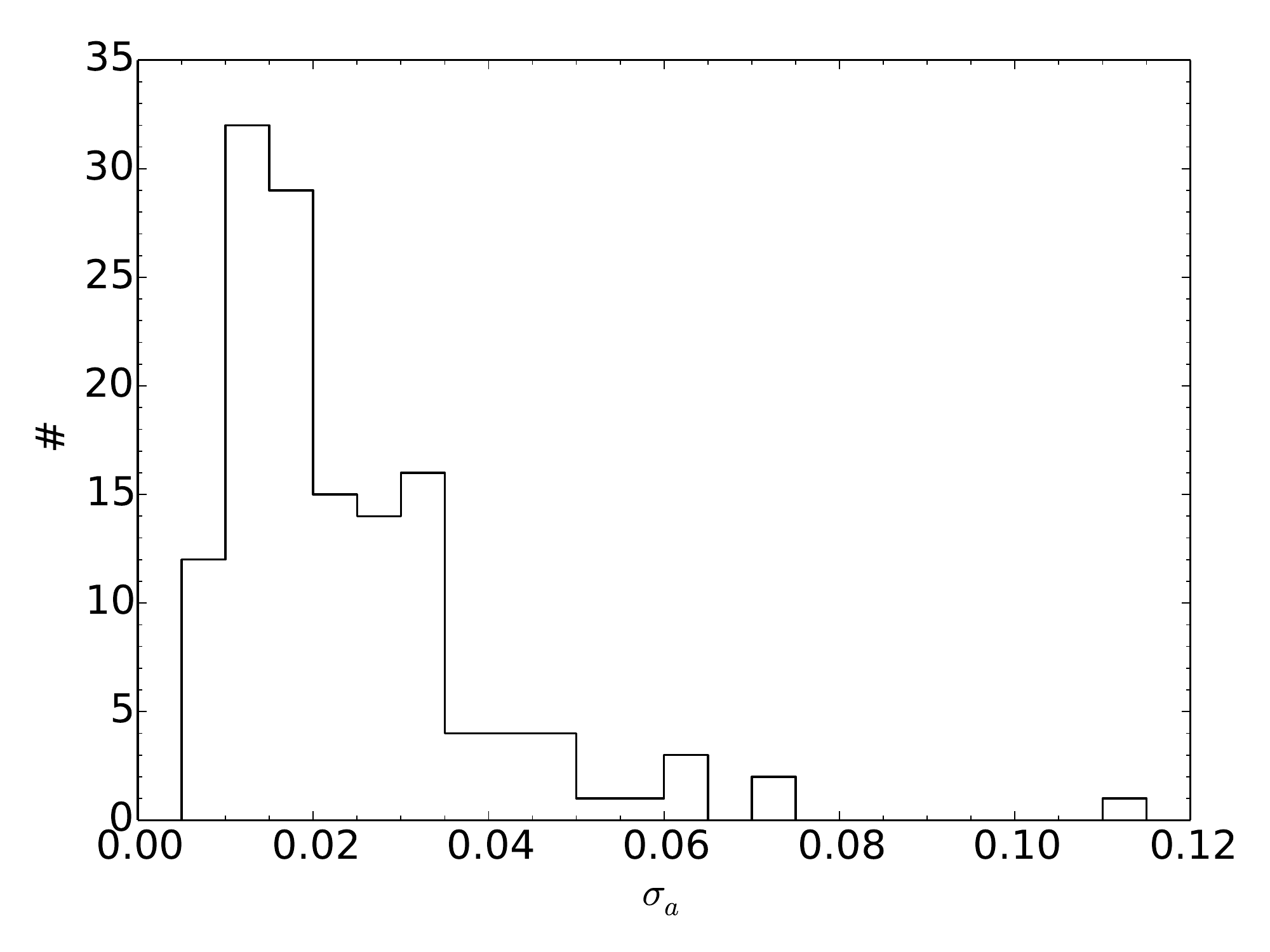}
\caption{Distribution of the uncertainty on the charge-excess fraction, $\sigma_{a}$, for individual air shower measurements binned in intervals of $0.005$ for all measured air showers.}
\label{fig:single_event_a_fit_uncertainty_distribution}
\end{figure}

\begin{figure}
\centering
\includegraphics[width=\figwidth\textwidth]{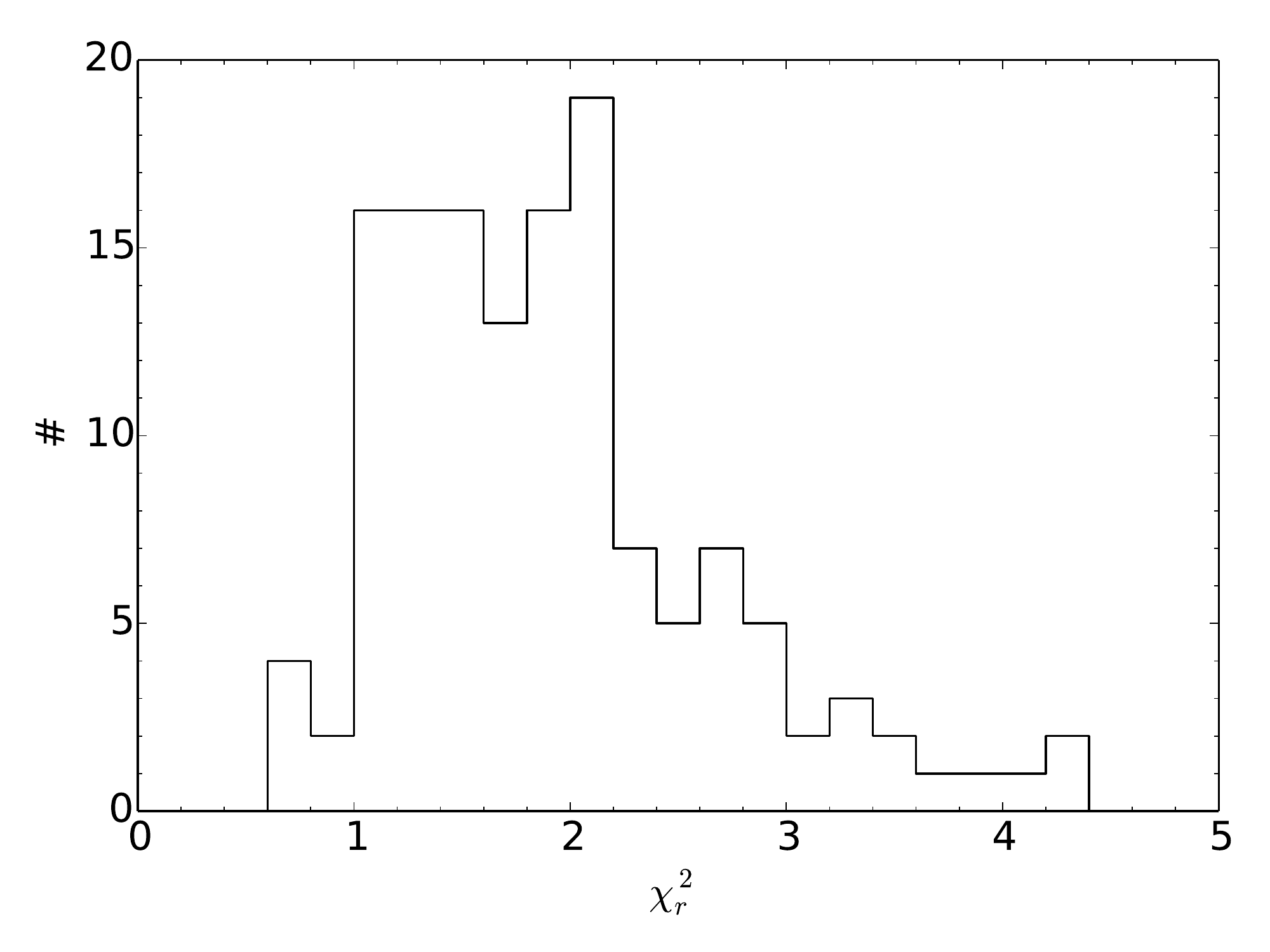}
\caption{Distribution of the fit quality, as parameterized by the reduced $\chi^2$ value, obtained when fitting the charge-excess fraction for individual air showers binned in intervals of $0.2$ for all measured air showers.}
\label{fig:chi2_single_event_a_fit}
\end{figure}

In order to minimize the influence of background noise only antennas with a measured signal-to-noise ratio exceeding three (in electric field amplitude) are included. Furthermore, a minimum of $48$ antennas (one full station) are required to ensure a stable fit result. The distribution of the best fitting values for the charge-excess fractions of all $138$ events surviving these cuts can be seen in figure~\ref{fig:distribution_single_event_a_fit}. These fits give an average charge-excess fraction of $(11 \pm 4)\%$ for our sample, where the uncertainty reflects the spread of the distribution. This measured presence of a radially polarized emission component is consistent with that produced by the charge-excess mechanism.
The uncertainty on the individual values of $a$ are determined as described in section~\ref{sec:uncertainty_charge_excess} and their distribution is plotted in figure~\ref{fig:single_event_a_fit_uncertainty_distribution}.
The fit quality, as parameterised by $\chi_{r}^{2}=\chi^2/\mathrm{dof}$, is given in figure~\ref{fig:chi2_single_event_a_fit}. The overall fit quality is reasonably good $\chi_{r}^{2}=1.9 \pm 0.7$ on average. However, theory predicts additional dependencies on the distance to the shower axis as well as the shower arrival direction \cite{de-Vries:2013}, these are not taken into account at this stage, which will necessarily lead to suboptimal fit results. For the same reason it is important to note that any quoted average charge-excess value will only apply to the specific set of showers for which it was measured. So while this analysis confirms that there is a radially polarized emission component, the average value itself has no meaning outside this sample. In the next section these dependencies are further investigated.

\subsection{Checking for additional dependencies on the geomagnetic angle}
\begin{figure}
\centering
\includegraphics[width=\figwidth\textwidth]{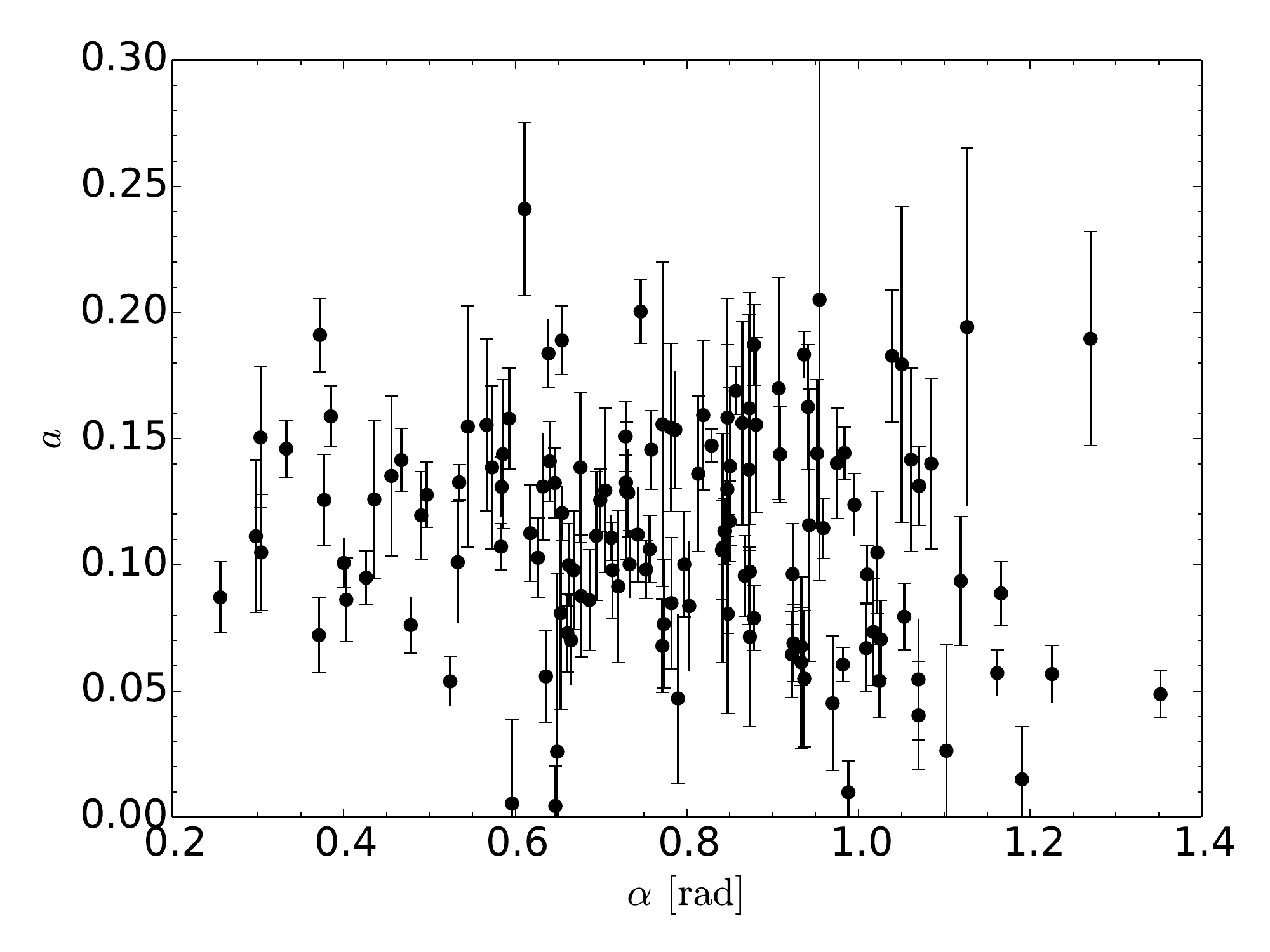}
\caption{Charge-excess fraction, $a$, as a function of the geomagnetic angle $\alpha$. Uncertainties are calculated as discussed in section~\ref{sec:uncertainty_charge_excess}.}
\label{fig:distribution-a-for-all-events}
\end{figure}

It is important to note that eq.~\eqref{eq:charge_excess_fraction} assumes that the charge-excess fraction $a$ only depends on the angle $\alpha$, that the propagation axis of the shower makes with the geomagnetic field, through the strength of the geomagnetic contribution which is proportional to $\sin\alpha$.
This assumption can now be checked by looking for an additional dependence of $a$ to $\alpha$ in figure~\ref{fig:distribution-a-for-all-events}, where the charge-excess fraction is plotted as a function of $\alpha$. We see no evidence for such a dependence. Note that the scatter of the points is greater than their uncertainties suggest. This indicates an additional dependence, that does not scale with the geomagnetic angle, which is almost certainly due to a dependence of the charge-excess fraction on the shower arrival direction and the distance to the shower axis. This additional dependence cannot be determined on a single air shower basis due to the limited number of data points available after radial and angular binning (see section~\ref{sec:distance_dependence}).

\subsection{Dependence on shower arrival direction and radial distance to the shower axis}
\label{sec:distance_dependence}
To verify the qualitative behavior predicted in \cite{de-Vries:2013}, an increase of the charge-excess fraction with increasing radial distance from the shower axis and decreasing zenith angle, all measured showers are combined in a single analysis. This gives a set of data points $\psi_i(\phi'_i, r'_i, \alpha_i,\theta_i)$ where $\alpha_i$ and $\theta_i$ are the geomagnetic angle and zenith angle of the corresponding air shower respectively. Subsequently these are grouped into radial distance bins of $\unit[50]{m}$ and zenith angle bins of $20^\circ$. For each group, $a$ was reconstructed by fitting eq.~\eqref{eq:polarization_angle_a} simultaneously to all $\psi_i$ values contained in it.

\begin{figure}
\centering
\includegraphics[width=\figwidth\textwidth]{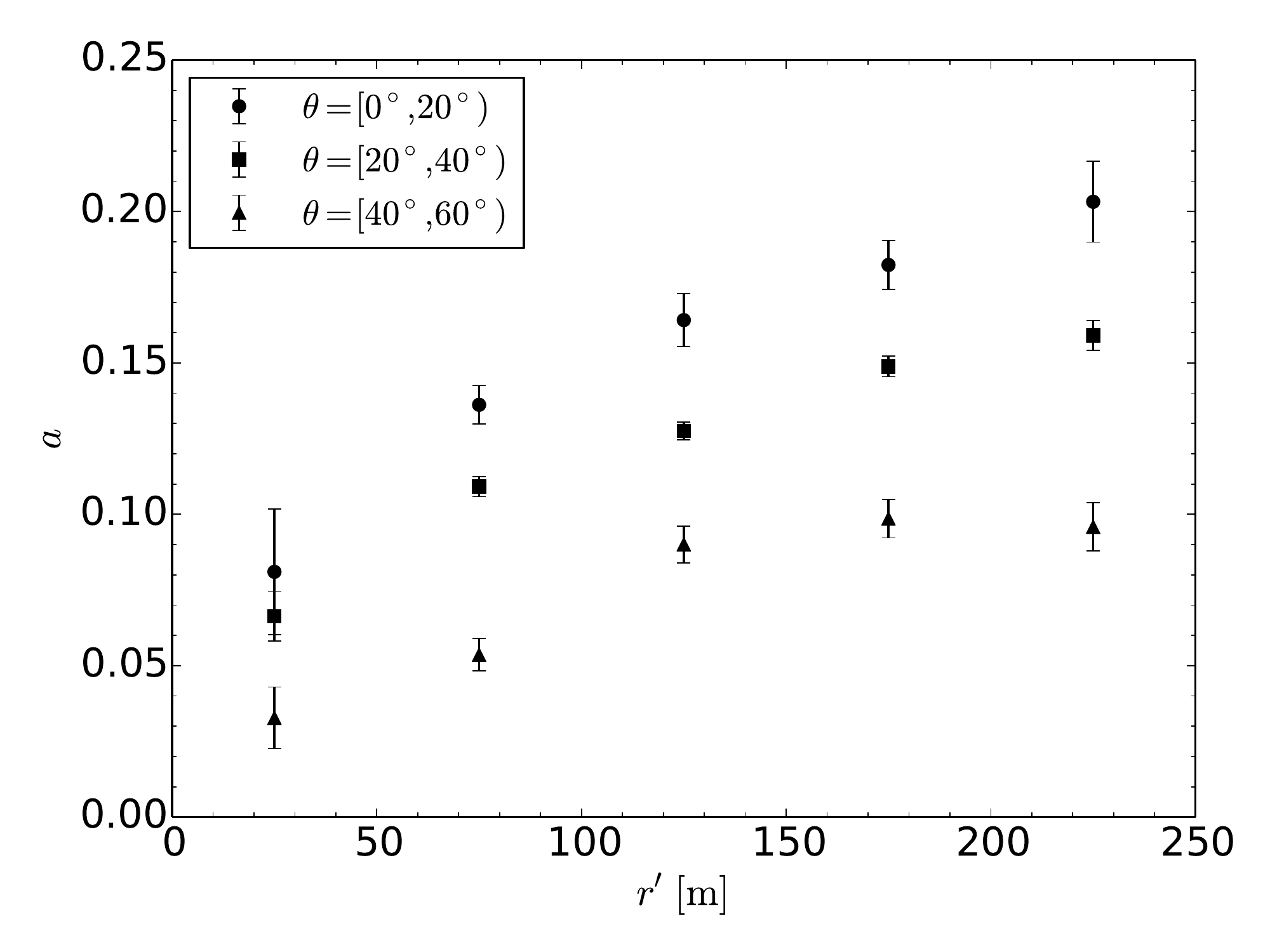}
\caption{Charge-excess fraction as a function of distance from the shower axis for three different zenith angle bins. Uncertainties are calculated as discussed in section~\ref{sec:uncertainty_charge_excess}.}
\label{fig:a-r-binned-50m-binned-zenith-20-deg}
\end{figure}

The result can be seen in table~\ref{tab:a-r-binned-50m-binned-zenith-20-deg} and in figure~\ref{fig:a-r-binned-50m-binned-zenith-20-deg}. The uncertainty on $a$ is determined as described in section~\ref{sec:uncertainty_charge_excess}.
\begin{table}
\caption{Charge-excess fraction as a function of the distance from the shower axis for three different zenith angle bins. Uncertainties are calculated as discussed in section~\ref{sec:uncertainty_charge_excess}.}
\label{tab:a-r-binned-50m-binned-zenith-20-deg}
\centering
\begin{tabular}{|c|c|c|c|}
\hline
     & \multicolumn{3}{|c|}{Charge-excess fraction ($a$)}\\
\hline
$r'$ & $\theta = [0^\circ, 20^\circ)$ & $\theta = [20^\circ, 40^\circ)$ & $\theta = [40^\circ, 60^\circ)$\\
\hline
$\unit[0-50]{m}$ & $(8.1\pm 2.1)\%$ & $(6.6\pm 0.8)\%$ & $(3.3\pm 1.0)\%$\\
$\unit[50-100]{m}$ & $(13.6\pm 0.6)\%$ & $(10.9\pm 0.3)\%$ & $(5.4\pm 0.5)\%$\\
$\unit[100-150]{m}$ & $(16.4\pm 0.9)\%$ & $(12.7\pm 0.3)\%$ & $(9.0\pm 0.6)\%$\\
$\unit[150-200]{m}$ & $(18.2\pm 0.8)\%$ & $(14.9\pm 0.3)\%$ & $(9.9\pm 0.6)\%$\\
$\unit[200-250]{m}$ & $(20.3\pm 1.3)\%$ & $(15.9\pm 0.5)\%$ & $(9.6\pm 0.8)\%$\\
\hline
\end{tabular}
\end{table}

Both the predicted increase with increasing radial distance, as well as the decrease with increasing zenith angle, can clearly be seen in figure~\ref{fig:a-r-binned-50m-binned-zenith-20-deg}. Note however that the specific values obtained, and listed in table~\ref{tab:a-r-binned-50m-binned-zenith-20-deg}, still depend on the event set due to shower-to-shower fluctuations.

\section{Discussion and conclusions}
\label{sec:conclusions}
Polarized radio emission from a sample of $163$ air showers measured with the LOFAR radio telescope has been analyzed.

In total $18$ air showers where excluded from this analysis due to coinciding thunderstorm activity. The strong atmospheric electric fields present during thunderstorms are expected to significantly affect the charged particle distributions and therefore the polarization of the emission \cite{Buitink:2007,Buitink:2010}. This effect will be investigated in a future publication.

The measured emission is strongly polarized, with a median degree of polarization of nearly $99\%$. Because the geomagnetic and charge-excess emission are linearly polarized in different directions, their relative contributions can be determined by polarization analysis.

In all measured air showers the geomagnetic emission mechanism clearly dominates the polarization pattern. However, a sub-dominant charge-excess component can also be seen, varying in strength between showers.

The relative strength of both contributions is quantified by the charge-excess fraction. We find that the measured charge-excess fraction is higher for air showers arriving from closer to the zenith. Furthermore, the measured charge-excess fraction also increases with increasing observer distance from the air shower symmetry axis. The measured values range from $(3.3\pm 1.0)\%$ for very inclined air showers at $\unit[25]{m}$ to $(20.3\pm 1.3)\%$ for almost vertical showers at $\unit[225]{m}$. Both dependencies are in qualitative agreement with theoretical predictions.

\acknowledgments
The LOFAR key science project Cosmic Rays very much acknowledges the scientific and technical support from ASTRON. Furthermore, we acknowledge financial support from the Netherlands Research School for Astronomy (NOVA), the Samenwerkingsverband Noord-Nederland (SNN), the Foundation for Fundamental Research on Matter (FOM) and the Netherlands Organization for Scientific Research (NWO), VENI grant 639-041-130. We acknowledge funding from an Advanced Grant of the European Research Council under the European Unions Seventh Framework Program (FP/2007-2013) / ERC Grant Agreement n. 227610.

LOFAR, the Low Frequency Array designed and constructed by ASTRON, has facilities in several countries, that are owned by various parties (each with their own funding sources), and that are collectively operated by the International LOFAR Telescope (ILT) foundation under a joint scientific policy.


\bibliographystyle{JHEP}
\bibliography{polarization.bib}

\end{document}